\documentclass[pdflatex,sn-mathphys-num]{sn-jnl}

\usepackage{graphicx}%
\usepackage{multirow}%
\usepackage{amsmath,amssymb,amsfonts}%
\usepackage{amsthm}%
\usepackage{mathrsfs}%
\usepackage[title]{appendix}%
\usepackage{xcolor}%
\usepackage{textcomp}%
\usepackage{manyfoot}%
\usepackage{booktabs}%
\usepackage{algorithm}%
\usepackage{algorithmicx}%
\usepackage{algpseudocode}%
\usepackage{listings}%

\theoremstyle{thmstyleone}%

\theoremstyle{thmstyletwo}%

\theoremstyle{thmstylethree}%

\begin{document}

\title[Exploring the Application of Quantum Technologies to Industrial and Real-World use cases]{Exploring the Application of Quantum Technologies to Industrial and Real-World use cases}

\author*[1]{\fnm{Eneko} \sur{Osaba}}\email{eneko.osaba@tecnalia.com}

\author[1]{\fnm{Esther} \sur{Villar-Rodriguez}}\email{esther.villar@tecnalia.com}

\author[1]{\fnm{Izaskun} \sur{Oregi}}\email{izaskun.oregui@tecnalia.com}

\affil[1]{\orgname{TECNALIA, Basque Research and Technology Alliance (BRTA)}, \orgaddress{\street{Geldo Auzoa, 700 Building}, \city{Derio}, \postcode{48160}, \country{Spain}}}

\abstract{Recent advancements in quantum computing are leading to an era of practical utility, enabling the tackling of increasingly complex problems. The goal of this era is to leverage quantum computing to solve real-world problems in fields such as machine learning, optimization, and material simulation, using revolutionary quantum methods and machines. All this progress has been achieved even while being immersed in the \textit{noisy intermediate-scale quantum} era, characterized by the current devices' inability to process medium-scale complex problems efficiently. Consequently, there has been a surge of interest in quantum algorithms in various fields. Multiple factors have played a role in this extraordinary development, with three being particularly noteworthy: \textit{i)} 
the development of larger devices with enhanced interconnections between their constituent qubits, \textit{ii)} the development of specialized frameworks, and \textit{iii)} the existence of well-known or ready-to-use hybrid schemes that simplify the method development process. In this context, this manuscript presents and overviews some recent contributions within this paradigm, showcasing the potential of quantum computing to emerge as a significant research catalyst in the fields of machine learning and optimization in the coming years.}


\maketitle

\section{Introduction} 

Quantum Computing (QC) marks a groundbreaking advancement in computational technology leveraging principles from quantum physics to handle data in entirely new manners \cite{memon2024quantum}. By exploiting quantum effects such as entanglement and superposition, purely quantum or hybrid algorithms are anticipated to offer significant improvements in speed and accuracy for system modeling and solving intricate problems.

Despite notable progress, quantum devices are still in their early stages compared to classical systems. Presently, they face difficulties related to the limited number of qubits and their instability. Problems like noise, information loss, and decoherence, particularly without error correction, negatively impact their performance. Furthermore, issues such as gate noise and quantum gate fidelity hinder advancements. Even hybrid algorithms have drawbacks; for instance, the physical separation of quantum and classical hardware introduces latency when they exchange information \cite{lubinski2022advancing}.

Consequently, we are now in the \textit{noisy intermediate-scale quantum} (NISQ, \cite{preskill2018quantum}) era, marked by the inefficiency of current devices in addressing complex problems. Despite these challenges, there has been a growing body of research in recent years focusing on addressing real-world problems using QC. The increasing volume of publications highlights the growing interest of the community in exploring QC applications. Several factors have contributed to this intriguing development:

\begin{itemize}
	\item \textit{The creation of larger and more interconnected devices}. Even within the NISQ era, systems as D-Wave's latest quantum annealer, which consists of 5616 qubits arranged in a Pegasus topology \cite{boothby2020next}, have enabled the tackling of more extensive problems. In fact, recent studies highlight the outstanding performance of such devices compared to classical solvers in what are called \textit{hardware-native problems} \cite{tasseff2024emerging}. Although these findings are preliminary, they suggest a promising future for D-Wave computers, which aspire to achieve \textit{quantum utility} in the coming years \cite{mcgeoch2022milestones}. Regarding IBM's gate-based computers, the Heron processor, featuring 156 superconducting qubits, marks a significant advancement also in this direction.
	\item  \textit{The development and enhancement of specialized frameworks}, designed to simplify the creation, implementation, and execution of quantum algorithms, have also influenced this growing interest. Noteworthy examples include Qiskit\footnote{\url{https://www.ibm.com/quantum/qiskit}}, Google's Cirq\footnote{\url{https://quantumai.google/cirq}} and NVIDIA's CUDA-Q\footnote{\url{https://developer.nvidia.com/cuda-q}}. These tools have fosterred a multidisciplinary community around quantum computing \cite{villar2023hybrid}, making it easier for professionals in fields like artificial intelligence and optimization to engage with quantum computing, even if they lack deep expertise in physics or quantum mechanics.
	
	\item \textit{The availability of well-known or ready-to-use hybrid methods that simplify the method development process}. Within the gate-based paradigm, various techniques can be considered, such as Variational Quantum Algorithms \cite{cerezo2021variational}. Additionally, D-Wave's \textit{Hybrid Solver Service} \cite{HSS}, which encompasses four distinct techniques tailored to address specific input categories and problem types, serves also as notable examples. 
\end{itemize}

Due to the increasing interest in this area, quantum computing has already been effectively utilized in numerous applications, with industrial problems being the most prominent examples. This editorial presents a special issue focused on the latest advancements in quantum computing, particularly from the perspective of its application to real-world problems. 

Out of the 93 different submissions received, only ten successfully passed the initial editorial review. Following a rigorous review of these ten manuscripts, five were ultimately accepted. These contributions encompass a broad spectrum of topics that are detailed throughout the rest of this article. We will conclude this editorial with a final note advocating for continued efforts in QC. This field is poised for an exciting future, replete with challenging research directions to be pursued.
 
\section{An Overview of this Special Issue}

The contributions included in this special issue collectively explore the transformative potential of quantum computing, with a special focus on quantum machine learning (QML) in various domains, including industrial applications, scientific research, and entertainment. The overarching theme is the significant advancements and practical applications of quantum technologies, which promise to revolutionize fields that require high computational power and complex data analysis. This special issue also explores the field of quantum-inspired evolutionary metaheuristics. Each article delves into specific use cases, demonstrating the potential of quantum algorithms in terms of accuracy, efficiency, and scalability. Furthermore, each study not only showcases the current capabilities of quantum computing but also outlines future directions and challenges. 

To begin with, Cisneros et al. present a novel approach in their work to predict rock block falls on slopes using QML algorithms. The research aims to improve the accuracy and efficiency of slope stability assessments, which are crucial to prevent rockfall disasters in mining operations. The methodology involves collecting geotechnical data from a mining site and using it to train both classical Machine Learning models and QML algorithms. The study introduces a new classification system called Rock Block Stability (RBS), which complements existing classifications like Rock Mass Rating (RMR) and Geological Strength Index (GSI). The RBS classification considers the orientation of slopes and discontinuities, providing a more comprehensive assessment of slope stability. The results show that QML algorithms outperform classical ML models in terms of accuracy, particularly in predicting true positives and true negatives. The study concludes that incorporating QML algorithms into slope stability evaluations represents a significant innovation, offering greater reliability and safety in mining operations.

In their contribution, Shahid et al. introduce a movie recommendation system that utilizes a Quantum Support Vector Machine (QSVM) to enhance the speed and accuracy of recommendations. The research addresses the limitations of traditional recommendation systems, such as long training times and high computational costs, by leveraging quantum algorithms. The methodology involves collecting and preprocessing data from movie review datasets, implementing classical SVM for baseline comparison, and encoding data for QSVM. The QSVM is executed using IBM's quantum computer, and its performance is compared with classical SVM. The results show that QSVM outperforms classical SVM, achieving a 96\% accuracy and an F1-score of 0.9693, compared to the classical SVM's 95.33\% accuracy and 0.9641 F1-score. The study highlights the potential of QSVM in handling complex datasets and improving the personalization of movie recommendations. Future research directions include optimizing quantum kernel functions and exploring the integration of QSVM with other quantum machine learning techniques.

Garate-Perez et al. explore the use of Variational Quantum Algorithms (VQAs) for solving phase field simulation problems in metal solidification in their paper. The focus is on modeling dendritic solidification, a crucial process to determine the mechanical properties of metals in additive manufacturing. The study compares a classical surrogate model based on extreme gradient boosting with a hybrid classical-quantum VQA model. The phase field model used in the study simulates the spatio-temporal evolution of solidification in metals, capturing the complex interplay of thermal, mechanical, and chemical phenomena. The methodology involves training surrogate models using a representative dataset derived from simulations. The results demonstrate that VQAs can serve as a viable alternative for computationally expensive surrogate modeling, offering a promising performance in accuracy and efficiency.

Qi et al. introduce a novel algorithm in their work that combines quantum particle swarm optimization (QPSO) with an extreme learning machine (ELM) to enhance intrusion detection systems. The authors propose a feature selection algorithm based on partitioned gains to reduce redundant features, thereby improving model training speed and accuracy. This article stands out as the only one in the collection that can be classified under the branch known as quantum-inspired metaheuristics. In brief, quantum-inspired algorithms are a specific type of evolutionary algorithms that operate based on concepts and principles of quantum computing, such as interference, coherence, and the qubit as the unit of information. However, as explained in \cite{zhang2011quantum}, these methods are designed for classical computers rather than quantum hardware. Thus, the QPSO-ELM algorithm is designed to achieve high training and detection speeds while maintaining high accuracy. Additionally, the paper presents a hidden layer node selection algorithm to optimize the trained model, reducing its size without compromising detection accuracy. Experimental results demonstrate that the proposed method outperforms existing baseline methods in terms of accuracy, precision, recall, and detection latency.

Finally, AbuGhanem traces the evolution of IBM's quantum computing technology in his paper, highlighting significant milestones and advancements. It begins with the development of early quantum processors like Canary and progresses to more advanced processors such as Eagle and Condor. The Condor processor, featuring 1,121 superconducting qubits, marks a significant achievement by surpassing the 1,000-qubit barrier. The article explores practical applications of quantum computing across various industries. In healthcare, for example, IBM's quantum systems enhance medical imaging and biomarker discovery, improving early detection and accuracy. In the banking sector, quantum computing aids in risk profiling and optimizing trading strategies. The article also addresses challenges related to scaling quantum systems and achieving fault tolerance. IBM's roadmap extends to 2033, focusing on developing modular quantum processors and quantum-centric supercomputing. Key milestones include the introduction of the Starling processor with 200 qubits and the Blue Jay processor with 2,000 error-corrected qubits, capable of performing 1 billion gates.

\section{Conclusion and Perspectives}

This special issue provides an inspiring overview of the current capabilities of quantum computing. The conclusions presented by the contributors should be seen not only as evidence of the vibrant research activity in quantum computing, machine learning, and optimization but also as an indication of the exciting future envisioned for this field.

This special issue provides an inspiring overview of the current capabilities of quantum computing. The conclusion presented by the contributors should be seen not only as evidence of the vibrant research activity in quantum computing, machine learning, and optimization but also as an indicator of the exciting future envisioned for this field. In summary, these articles collectively paint a picture of the quantum revolution, illustrating how this field is set to redefine industries, drive scientific breakthroughs, and enhance everyday applications. 

Moreover, the advancement of technology continually provides us with the opportunity to work with more advanced and reliable devices, which, as noted in papers as \cite{abbas2024challenges} or \cite{cerezo2022challenges}, have immense future potential. 

Furthermore, the development of scalable quantum systems and the achievement of fault tolerance are critical milestones that researchers and industry leaders are working toward, contributing to the anticipation of a promising future for the field. IBM's roadmap, for example, envisions the creation of modular quantum processors and quantum-centric supercomputing by 2033, with significant advancements expected along the way. Another notable example is the quantum annealers developed by D-Wave, which are projected to achieve 100,000 qubits by the end of this decade, as outlined in the recently unveiled roadmap by the Canadian company\footnote{\url{https://quantumcomputingreport.com/recap-d-wave-qubits-2025-conference/}}.

Given the recent advancements, which the special issue highlighted in this editorial strongly attests to, it is indisputable that the opportunities and benefits arising from the evolution of quantum computing will soon surpass the boundaries of our imagination.

\section*{Acknowledgements}

This work was supported by the Basque Government through Plan complementario comunicación cuántica (EXP. 2022/01341) (A/20220551). During the preparation of this work, the authors used Microsoft Copilot to improve the language and readability of the manuscript. After using this tool/service, the authors reviewed and edited the content as needed, taking full responsibility for the content of the publication.

\bibliography{sn-bibliography}

\end{document}